\documentclass[aps,prl,twocolumn,showpacs,a4paper,superscriptaddress]{revtex4}
\usepackage{graphicx}
\usepackage{amsmath}
\begin{document}
\title{Electron states of mono- and bilayer graphene on SiC probed by STM}
\author{P. Mallet}
\affiliation{Institut N\'eel, CNRS - Universit\'e Joseph Fourier, BP166, F-38042 Grenoble Cedex 9, France\\}

\author{F. Varchon}
\affiliation{Institut N\'eel, CNRS - Universit\'e Joseph Fourier, BP166, F-38042 Grenoble Cedex 9, France\\}

\author{ C. Naud}
 \affiliation{Institut N\'eel, CNRS - Universit\'e Joseph Fourier, BP166, F-38042 Grenoble Cedex 9, France\\}

\author{L. Magaud}
\affiliation{Institut N\'eel, CNRS - Universit\'e Joseph Fourier, BP166, F-38042 Grenoble Cedex 9, France\\}

\author{C. Berger}
\affiliation{Institut N\'eel, CNRS - Universit\'e Joseph Fourier, BP166, F-38042 Grenoble Cedex 9, France\\}
\affiliation{Georgia Institute of Technology, Atlanta, Georgia
30332-0430, USA\\}

\author{J.-Y.Veuillen}
\affiliation{Institut N\'eel, CNRS - Universit\'e Joseph Fourier, BP166, F-38042 Grenoble Cedex 9, France\\}

\date{\today}

\begin{abstract}
We present a scanning tunneling microscopy (STM) study of a gently-graphitized 6H-SiC(0001) surface in ultra high vacuum. From an analysis of atomic scale images, we identify two different kinds of terraces, which we unambiguously attribute to mono- and bilayer graphene capping a C-rich interface. At low temperature, both terraces show $(\sqrt{3}\times \sqrt{3})$ quantum interferences generated by static impurities. Such interferences are a fingerprint of $\pi$-like states close to the Fermi level. We conclude that the metallic states of the first graphene layer are almost unperturbed by the underlying interface, in agreement with recent photoemission experiments (A. Bostwick et al., Nature Physics 3, 36 (2007)).

\end{abstract}

\pacs{73.20.-r, 68.37.Ef, 72.10.Fk}




\maketitle

Although the first band structure calculation of graphene (one sp$^{2}$ bonded carbon layer) has been performed almost 60 years ago \cite{Wallace}, the experimental proof of the remarkable electron properties of this system has been reported only recently. In particular, the predicted Dirac character of graphene fermions close to the Fermi level ($E_{F}$) has been shown, giving rise to anomalous integer quantum hall effect and phase shifted Shubnikov de Hass oscillations \cite{Novoselov,Zhang,Berger}. For these pioneering experiments, ingenious techniques were applied to isolate graphene layers, either by graphite exfoliation \cite{Novoselov,Zhang} or by graphitization of SiC \cite{Forbeaux,Berger}.

For both methods, decoupling of the graphene wave functions from the neighboring environment is a fundamental issue. In graphitized SiC surfaces, the graphene layer(s) is (are) separated from the bulk  by a carbon rich interlayer which is of primary importance.
 Very recently, Angle-resolved Photoemission Spectroscopy (ARPES) measurements were reported on both bilayer and monolayer graphene obtained on a graphitized n-type doped SiC(0001) substrate \cite{Bostwick,Ohta}. For the graphene monolayer, the Dirac-like character of carriers was evidenced from the linear dispersion close to the Dirac point (the  point were hole and electron bands touch each other), and many-body interactions in this two-dimensional (2D) system could be studied \cite{Bostwick}. This is a clear demonstration that the C rich interface has an almost negligible influence on the surface Dirac-like carriers, as previously suggested \cite{Forbeaux, Berger, Magaud}. Apart from electron doping of the graphene layer due to charge transfer from the bulk, the conduction states of this system can be considered as those of an almost free-standing graphene sheet  \cite{Bostwick}.

    Scanning Tunneling Microscopy (STM) is a powerful technique for studying surface (quasi) 2D states at the atomic scale  \cite{Eigler,Brihuega,Ono}. However, no direct STM investigation of the graphene low energy states has been reported. This technique has been used to characterize the surface morphology down to atomic scale for different stages of the graphitization of SiC(0001) \cite{Owman,Charrier,Chen,Berger}.  Interestingly high bias images of areas with one graphene monolayer shown in references \cite{Owman,Charrier,Berger} are often dominated by a strong contrast related to the interface. This might be interpreted as an evidence of a strong interaction between interface states and surface states, in apparent contradiction with ARPES results of Ref. \cite{Bostwick}. A second issue is related to the exact number of graphene layers below the STM tip. In particular, a clear fingerprint of the single graphene layer has not been demonstrated. This point must be overstepped for future STM investigations of the unique electron properties of graphene.

In this letter, we present an STM study of the initial stages of graphitization of a 6H-SiC(0001) substrate. Starting with the precursor C rich phase, the so-called  $(6\sqrt{3}\times6\sqrt{3})$ reconstruction, the sample is annealed to promote the synthesis of a few graphene layers. The surface becomes metallic, as shown by low bias STM images at $T$ = 45K which are routinely achieved. From the analysis of the STM contrast at atomic scale, two different phases are identified, which are attributed to single and double graphene layer. For both phases, quantum interferences are found in the vicinity of impurities, leading to a $(\sqrt{3}\times \sqrt{3})R30^{\circ}$ superstructure with respect to the graphene (1$\times$1) lattice. Such interferences originate from intervalley coupling of graphene (graphite) $\pi$-like states. Our atomic scale investigation demonstrates that the metallic states of the single graphene layer are essentially not affected by underlying interface states.

The sample preparation was done in ultra high vacuum, with low energy electron diffraction (LEED) and Auger spectroscopy, following the procedure of Forbeaux et al \cite{Forbeaux}. A n-type (nitrogen 1 $\times$ 10$^{18}$ cm$^{-3}$)  6H-SiC(0001) (i.e. Si terminated) substrate was first heated at 900$^{\circ}$C under a low Si flux, producing a ($3\times3$) Si rich phase.
Successive annealings at increasing temperatures (from 900 to 1100$^{\circ}$C) led first to a $(\sqrt{3}\times \sqrt{3})R30^{\circ}$ (R3) phase, then to a C-rich phase with a $(6\sqrt{3}\times 6\sqrt{3})R30^{\circ}$ (6R3) reconstruction.
As reported in \cite{Owman,Chen}, R3 spots, which initially coexist with the 6R3 spots in the LEED pattern, disappear with further annealings. The pattern shown in Fig. 1 (a), obtained with a primary energy of 109 eV, exhibits SiC(0001) (1$\times$1) spots surrounded by hexagonal 6$\times$6 spots (SiC-6$\times$6 in the following), and also faint 6R3 spots.

An STM image of this surface, recorded at 45K and at sample bias -2.0 V, is shown in Fig. 1 (b). It is similar to occupied states-images of the carbon nanomesh of Ref. \cite{Chen}, and to some images of  Ref. \cite{Owman,Charrier}. The honeycomb structure close to a SiC-6$\times$6 is related to the 6R3 reconstruction. It is always observed and has been largely discussed in these references. However, the precise atomic structure and the related electron properties of the actual reconstruction are far from fully understood.  As shown in Ref. \cite{Chen}, the large corrugation found on the terraces is not of electronic origin. The authors have suggested that the whole surface is covered by tiny graphene-like carbon islands, self-organized to form the honeycomb structures, with part of C atoms forming covalent bonds with Si atoms \cite{Chen}. However no atomic resolution has been achieved on this surface to our knowledge, so that there is no direct evidence of such local graphene-like structure.  Additionally, ARPES measurements have identified $\sigma$-bands related to graphitic $sp^{2}$ bonded carbon, but have pointed out the lack of $\pi$-like bands in the vicinity of $E_{F}$ \cite{Emtsev}. We note that STM images at low bias are not achievable, neither at room temperature nor at 45K. This points to a non metallic character of the surface (the substrate is insulating at 45K), which implies that the first C rich layer has no graphene-like electron properties close to $E_{F}$.

\begin{figure}
\includegraphics[width=7.5cm,clip]{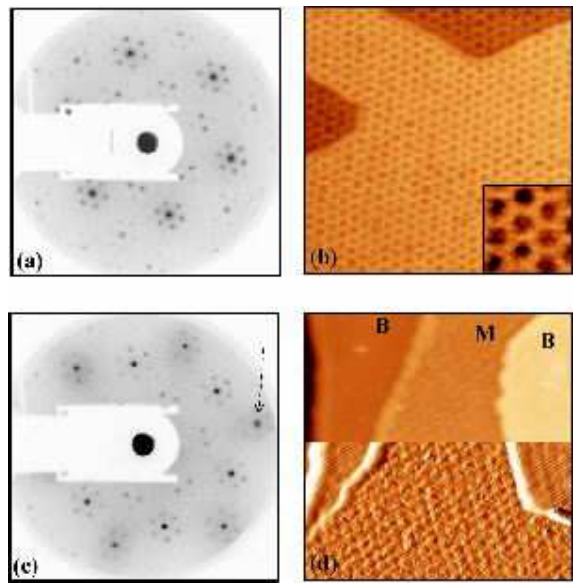}
\caption{(color online) (a) 109 eV LEED pattern of the 6H-SiC(0001) 6R3 reconstruction. (b) 40 $\times$ 40 nm$^{2}$ STM image at 45 K of the same surface, exhibiting the carbon nanomesh phase \cite{Chen}. Sample bias: -2.0 V.  Inset: a 5$\times$5 nm$^{2}$ zoom of the central terrace. (c,d) equivalent data of (a,b) after the last annealing step. (c) the dashed arrow indicates one of the graphene (1$\times$1) spots. (d)  The bottom part of the image is derivated to highlight the different corrugation between terraces M and B. Sample bias: +0.5 V. }
\label{f.1}
\end{figure}

 In the following, we study the same sample after a subsequent annealing at 1300-1350$^{\circ}$C during 8 mn. The C:Si Auger ratio does not exceed 2, indicating that only a few C layers are present onto the surface. As shown in  Fig. 1 (c), the surface layers have the lattice periodicity of a graphene sheet: pronounced spots of the (1$\times$1) graphene lattice are found in the LEED pattern, in addition with the SiC related spots. STM images of large areas, such as Fig. 1 (d), reveal terraces with a periodic superstructure, the lattice parameter of which corresponds to that of the SiC-6$\times$6 shown in Fig. 1 (b).  Therefore, this superstructure is induced by the C-rich interface lying just below the graphene sheets. Surprisingly, we find that the corrugation of this superstructure is not the same for all terraces. This is demonstrated in the lower part of Fig. 1 (d), where the image has been derivated. The central terrace (labelled M) exhibits a higher corrugation than the other terraces (labelled B), with a factor between 3 to 5 depending on the sample bias. We have checked that most of the terraces studied on this sample are either of M or B type (their identification is made easy on large scale derivated images).

 To elucidate the nature of the two different terraces, we focus on STM images with atomic resolution. Such images are routinely achieved at 45K with sample bias as low as 50 mV, which means that the surface is metallic. Fig. 2 is a panel of typical STM images, at sample bias + 0.2V. For type M terraces shown in Fig. 2 (a-c), images reveal a graphene (1$\times$1) lattice of dark spots (with a measured lattice parameter of 2.4 $\pm$ 0.2 ${\AA}$). The six C atoms surrounding each spot give the same bright signal, which leads to a true honeycomb atomic pattern (symmetric contrast). As quoted above, images of type M terraces such as Fig. 2 (a) are also frequently dominated by features related to the C-rich (6$\times$6) interface, which are superimposed to the graphene (1$\times$1) pattern (see also Fig. 3 (c)). Occasionally, uncontrolled change of the tip apex gives rise to a strong attenuation of this interface contribution. This is illustrated on Fig. 2 (b), where only a smooth SiC-6$\times$6 pattern remains together with the graphene (1$\times$1) lattice. Fig 2 (a) and (b) correspond to the same area and the contrast difference between the two images is only due to a tip apex modification. The honeycomb atomic pattern is not affected by this tip effect.

  \begin{figure}
\includegraphics[width=6.5cm,clip]{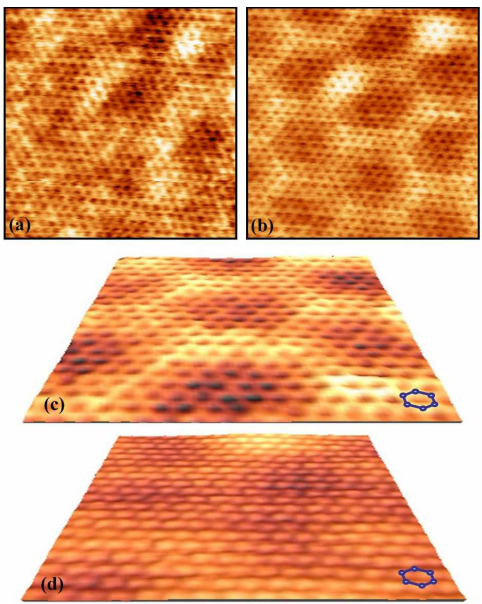}
\caption{(color online) (a-b) 5.6 $\times$ 5.6 nm$^{2}$ STM images of the same area of an M type terrace, with an unexpected tip change between the two images.  (c-d) 4 $\times$ 4 nm$^{2}$ 3D view of an M-monolayer (c) and a B-bilayer graphene (d) terrace. A hexagonal graphene unit-cell is depicted on both images. Sample bias was set at +0.2V for all the images.}
\label{f.2}
\end{figure}

We compare now a 3D zoom of Fig. 2 (b), shown in Fig. 2 (c), with the equivalent data for a B type terrace (Fig. 2 (d)). Two differences are found between the images. First, the SiC-6$\times$6 superstructure is weaker for terraces B than terraces M. Second, and more important, the atomic pattern observed on B terraces shows an asymmetric contrast: bright spots originate from only three C atoms out of six of a graphene unit cell. This asymmetric contrast is commonly reported for highly oriented pyrolitic graphite (HOPG) surfaces \cite{Hembacher}.
The results concerning the symmetric (asymmetric) atomic contrast found on M (B) type terraces are general and robust. Occasionally, unexpected tip modification may lead to puzzling contrasts, as reported for HOPG \cite{Atamny}.

Results of Fig. 2 can be interpreted in a very simple manner by attributing type M terraces to monolayer graphene covering the C rich interface. A symmetric atomic contrast has been reported recently for a graphene monolayer on Ir(111) \cite{grapheneIr}. This is intuitively expected for one single graphene layer that is essentially decoupled from the substrate, since all C atoms of the layer are equivalent. On the other hand, surface C atoms of a graphene bilayer with AB stacking become inequivalent, as in HOPG. The B terrace corresponds to bilayer graphene, since the graphitization of the SiC(0001) surface is a layer by layer process \cite{Charrier}.

We attribute terrace M to one graphene layer also from the analysis of the contribution of the C-rich interface layer to the STM images. As seen above, the corresponding corrugation is weaker for terraces B than M type. Having a closer look to images of M terraces, we find that atomic details of the interface can often be distinguished "through" the honeycomb atomic pattern (Fig. 2 (a) and 3 (c)). This observation can also be found in previous reports on graphitized SiC \cite{Owman,Charrier,Berger}, on areas attributed to only one single graphene layer. Our interpretation of such contrast is the following: for one graphene monolayer and at low bias, we expect the tip to probe graphene metallic states but also possible states located just below the surface, namely at the C-rich interface. This is indeed possible due to the peculiar shape of the graphene Fermi surface, where only high momentum 2D states exist. In that case, tunneling between the tip and the interface will occur for electrons having wave vector with small parallel component $k_{//}$, through the graphene layer which has no small $k_{//}$ available.
 This tunneling process is hindered in the case of a graphene bilayer, because of the increased tip-interface distance ($\sim$3.5 ${\AA}$, i.e. a graphite interlayer distance).

Our interpretation for the strong interface STM contrast on terraces M requires interface states (below the single graphene layer) close to $E_{F}$. From ARPES \cite{Bostwick} and Momentum-resolved Inverse Photoemission Spectroscopy (KRIPES) \cite{Forbeaux}, it appears that the $\pi$-like bands of the graphene monolayer on 6H-SiC(0001) are essentially not affected by any interface states.  Recent density functional theory calculations show that remaining dangling bonds of the complex carbon interface give rise to interface states, which however preserve the Dirac dispersion of the first graphene layer \cite{Magaud}. Confirmation of a weak interaction between such interface states and the metallic states of the surface is also shown on Fig. 2 (c), in which the graphene lattice appears almost perfect at the atomic scale. The remaining tiny SiC-6$\times$6 modulation is probably a real deformation of the surface layer, and apparently has no incidence on the surface electron properties close to $E_{F}$ \cite{Bostwick}.

 In the last part of this letter, we focus on the character of the metallic states probed either on M or B terraces. For that purpose,
 we use the STM tip as a local probe of the Local Density of States (LDOS) at the vicinity of static defects. Some impurities (of unknown nature) are located on top of the surface (they eventually can be swept by the STM tip).
 An impurity on a B type terrace (graphene bilayer) is shown on Fig. 3 (a). The sample voltage was fixed at -0.1 V. Superimposed to the (1$\times$1) atomic lattice, a $(\sqrt{3}\times \sqrt{3})R30^{\circ}$ (R3) superstructure surrounds the impurity, with a lateral extension of $\sim$ 5 nm. The corresponding FFT is shown on Fig. 3 (b), exhibiting the (1$\times$1) and the R3 spots. The R3 superstructure is commonly found at many impurities of B type terraces, for positive or negative sample bias as low as 10 mV. For M type terraces (graphene monolayer), the R3 superstructure is much more difficult to identify. The main reason is that only very few effective impurities (i.e. generating R3 superstructure) are identified on M terraces. Moreover, their observations are made difficult due to the strong corrugation generated by the interface. Image of Fig. 3 (c), recorded at sample bias -0.5V, shows evidence of such an R3 superstructure, but the impurity generating this pattern does not appear clearly on the image. The R3 pattern is also found on low bias images for this terrace (not shown). Importantly, interface states dominating the contrast on most images of terraces M do not induce any R3 pattern (see. Fig 2. (a) and 3. (c)), which supports once again an efficient decoupling of the graphene layer.

\begin{figure}
\includegraphics[width=7.5cm,clip]{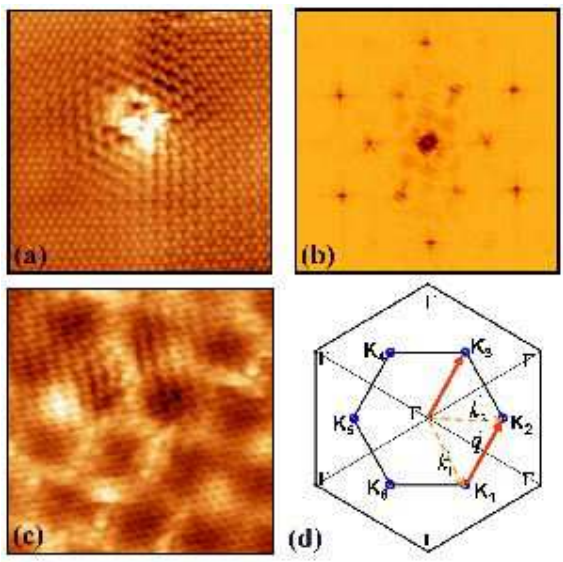}
\caption{(color online) (a) Impurity-induced quantum interferences on bilayer graphene. Image size: 7 $\times$7 nm$^{2}$, sample bias: -0.1V. (b) FFT of (a). outer spots: (1x1) atomic lattice, inner spots: R3 superstructure. (c) QI on monolayer graphene. Image size 7 $\times$7 nm$^{2}$, sample bias: -0.5V  (d) Illustration of intervalley coupling for graphene $\pi$ states at the Fermi level. Figures (a) and (c) were obtained at $T$ = 45K.}
\label{f.2}
\end{figure}

    The R3 superstructure around impurities has been reported for HOPG graphite surfaces \cite{Mizes,Ruffieux}, and also for one single graphene layer on Ir(111) \cite{R3grapheneIr}. This pattern is related to quantum interferences (QI) of $\pi$-like states scattered by an impurity, as illustrated in Fig. 3 (d). We have plotted a schematic Fermi surface (FS) of a lightly n-doped graphene monolayer (this picture is also valid for graphene bilayer). The FS consists in circular tiny pockets around K symmetry points of the graphene Brillouin zone. Scattering by an impurity between a state $\overrightarrow{k}_{1}$ and a state $\overrightarrow{k}_{2}$ of two adjacent pockets leads to LDOS spatial modulation with wave vector $\overrightarrow{q}=\overrightarrow{k}_{2}-\overrightarrow{k}_{1}$, i.e. $\overrightarrow{q}\simeq \overrightarrow{\Gamma K}_{3}$ for states depicted in Fig. 3 (d). The hexagonal symmetry of the FS leads to the R3 modulation in the LDOS, which is essentially recovered in constant current STM images of Fig. 3 (a) and 3 (c) \cite{Note}.

    The observation of the R3 pattern at impurities of terraces M or B demonstrates that the STM tip probes graphene $\pi$-like states on the surface, and that such states are essentially not altered by interface or substrate states, as shown by ARPES \cite{Bostwick,Ohta}. To our knowledge, this is the first report of R3 QI on a single graphene monolayer on an insulating substrate. As seen above, the R3 pattern is a proof of intervalley scattering, which is a key issue for transport properties in graphene. In particular, it should play a role in the quantum corrections to the electrical conductivity, with subtle effects since adjacent valley are nonequivalent in graphene \cite{Suzuura,McCann}.

       In conclusion, we studied the local electron properties of graphene monolayer and bilayer grown on a SiC substrate. STM allows a clear identification of the two systems and confirms the effective electron decoupling between the graphene layers and the substrate. Furthermore, STM offers the opportunity to probe scattering processes at impurities, consistent with the expected shape of the mono- and bilayer graphene Fermi surface.

The authors thank Didier Mayou, Valerio Olevano, Laurent L\'evy and W.A. de Heer for very fruitful discussions.






\begin{thebibliography}{99}
\bibitem{Note}
Constant current images at low bias can be approximated to images of the LDOS at $E_{F}$, given our limited instrumental resolution and the very high dispersion of the $\pi$ band reported in Ref. \cite{Bostwick,Ohta}.
\end{thebibliography}
\end{document}